\documentclass[prd,preprint,tightenlines,floatfix,showpacs,preprintnumbers,nofootinbib,eqsecnum]{revtex4}
 \usepackage[dvips,final]{graphicx}
  \usepackage{amssymb}
   \usepackage{amsmath}
    \usepackage{amsfonts}
     \usepackage{epsfig}
      \usepackage{bm}

\begin{document}

\title{Complex-mass definition\\ and the structure of unstable particle's propagator}

\author{Vladimir Kuksa}
\affiliation{Research Institute of Physics, Southern Federal
University, 344090 Rostov-on-Don, Pr. Stachky 194, Russian
Federation\vspace{1cm}\footnote{vkuksa47@mail.ru}}

\begin{abstract}
The propagators of unstable particles are considered in
framework of the convolution representation. Spectral function is
found for a special case when the propagator of scalar unstable
particle has Breit-Wigner form. The expressions for the dressed
propagators of unstable vector and spinor fields are derived in an
analytical way for this case. We obtain the propagators in modified
Breit-Wigner forms which correspond to the complex-mass definition.
\end{abstract}


\maketitle

\section{Introduction}

 Two standard definitions of the mass and
width of unstable particles (UP), which are usually considered in
the literature, have different nature. The on-mass-shell (OMS)
scheme defines the mass $M$ and width $\Gamma$ of UP by the
renormalization of the self-energy amplitude. In the pole scheme
(PS) the definitions of mass and width are based on the
complex-valued position of the propagator pole
$s_R-M^2_0-\Pi(s_R)=0$. There has been considerable discussion
concerning definition of the vector-boson mass
\cite{Sirlin,Willenb,LopezC,Stuart,Leike,Kniehl,Lucio,Faist,Bohm0,Zhou,Cacc}.
It was shown that OMS scheme contains spurious higher-order
gauge-dependent terms. Moreover, at one loop in the conventional OMS
the problem of threshold singularity arises which originates from
the wave-function renormalization constant $Z^{-1}=1-\Re A^{'}(M^2)$
\cite{Kniehl}. The PS provides gauge invariant definition and make
it possible to solve the problem of threshold singularity
\cite{Kniehl, Lucio}. However, it does not define the mass and width
in uniquely way \cite{Bohm}. One of the PS definition, where mass
$M_{\rho}$ and width $\Gamma_{\rho}$ are follows from the
parametrization $s_R=M^2_{\rho}-iM_{\rho}\Gamma_{\rho}$ \cite{Bohm},
is known as complex-mass definition. It should be noted that the PS
definition of the mass and width is connected with the structure of
the dressed propagator. Some aspects of the above mentioned problems
are considered further in more detail.

 Traditional way to construct the dressed propagator of UP is
Dyson summation which introduces the width and redefines the mass of
UP. This procedure runs into some problems which are widely
discussed in the literature. One of such problems follows from the
d'Alembert convergence criterion $|z|<1$ of the series
\begin{equation}\notag
\frac{1}{1-z}=\sum_{k=0}^{\infty}z^k=1+z+z^2+...,\qquad |z|<1,
\end{equation}
where $z=\Pi_{(1)}(q)/(q^2-M^2_0)$ and $\Pi_{(1)}(q)$ is the
one-particle--irreducible self-energy. The variable $z$ should be
correctly redefined before summation, that is we have to perform the
renormalization of the $\Pi_{(1)}(q)$ at Lagrangian level. This
procedure must be consistent with the infinite Dyson summation and
we can not use it after the redefinition at $|z|>1$. For instance,
in the resonant-part approximation we can define
$z=\Im\Pi_{(1)}(q)/(q^2-M^2)\approx M\Gamma/(q^2-M^2)$. So, the peak
range $|q^2-M^2|<M\Gamma$ or $|q-M|<\Gamma/2$ (at $\Gamma\ll M$) is
excluded by d'Alembert convergence criterion. There are, also, the
difficulties in the scheme of sequential fixed-order calculations
which exhibit themselves in the violation of the gauge invariance.
Moreover, using different decompositions of self-energy tensor in
the Dyson summation leads to different expressions for vector
dressed propagator \cite{Nowak,LopezC2,Liu}. Then, the
renormalization procedure is connected with the truncation of a
Laurent series expansion at the resonance range. So, the
renormalized propagator is an approximation of the full one which
corresponds to exact two-point function.

The peculiarities of Dyson summation lead to the lack of uniqueness
in constructing the propagators of unstable particles. There are
several different expressions for the numerator of vector boson
propagator $g_{\mu\nu}-q_{\mu}q_{\nu}/f(q,M,\Gamma)$, which are
exploited in practical calculations and give almost the same
numerical results. The denominator $f(q,M,\Gamma)$ has the following
forms (in the unitary gauge): $M^2,\, M^2-iM\Gamma,\,
(M-i\Gamma/2)^2,\, q^2$ and other combinations with $q$ -dependent
$M, \Gamma$. So, we need in an additional argumentation concern
these semiphenomenological definitions. It is known that the
commonly used Breit-Wigner (BW) expressions for bosonic and
fermionic propagators, respectively
\begin{equation}\label{1.1}
D^V_{\mu\nu}(q^2)=\frac{-g_{\mu\nu}+q_{\mu}q_{\nu}/M^2_V}{q^2-M^2_V+iM_V\Gamma_V};
\,\,\,D_F(\hat{q})=\frac{\hat{q}+M_F}{q^2-M^2_F+iM_F\Gamma_F}
\end{equation}
do not satisfy the electromagnetic Ward identity \cite{LopezC}. It
was shown in Refs.\cite{LopezC,Nowak,LopezC2}, that the modified BW
propagators
\begin{equation}\label{1.2}
D^V_{\mu\nu}(q^2)=\frac{-g_{\mu\nu}+q_{\mu}q_{\nu}/(M^2_V-iM_V\Gamma_V)}{q^2-(M^2_V-iM_V\Gamma_V)};
\,\,\,D_F(\hat{q})=\frac{\hat{q}+M_F-i\Gamma_F/2}{q^2-(M_F-i\Gamma_F/2)^2}.
\end{equation}
satisfy the electromagnetic Ward identity which provides the gauge
invariant description of processes with UP participation. Note, here
we deal with the resonant part of the full propagator which follows
from Dyson summation. The correctness of this propagator should be
understood in the context of resonant processes where the usage of
the modified BW propagator leads consistently to gauge invariant
results \cite{LopezC,Nowak,LopezC2}. It was also noted in Ref.
\cite{Nowak}, that in this case, we have to make the modification
$M^2_V\to M^2_V-iM_V\Gamma_V$ not only in the $q_{\mu}q_{\nu}$ term
of the propagator, but in the vertexes too. Thus, we get the
so-called complex-mass definition which was developed in the
framework of the complex-mass scheme (CMS)
\cite{Denner1,Denner2,Denner3,Djukan,Kniehl2}. Recently, BW
parametrization of the resonance lines has been developed in the
works \cite{Cruz,Ceci,Ceci1}. In particular, a new parameter has
been introduced into this parametrization which describes a
fundamental property of a resonance \cite{Ceci1}.

All above mentioned definitions are connected with the structure of
dressed propagators which follows from Dyson summation. As was noted
above, this procedure runs into some problems which are widely
discussed in literature. An alternative approach is based on the
spectral representation of the propagator of UP. It has a long
history \cite{Lehm, Gellman, Kallen, Matth, Schwing, Jacob, Greenb,
Lich,Burnier, Dudal, Lutz} and treats UP as a non-perturbative state
or effective field (asymptotic free field \cite{Greenb, Lich}). For
the first time, the hypothesis of continuous (smeared) mass of UP
was suggested by Matthews and Salam \cite{Matth}. In this paper,
they have formulated ``...a very direct interpretation to the
spectral function introduced by K\"{a}ll\'{e}n \cite{Kallen} and
Lehmann \cite{Lehm}''. The authors interpreted the spectral function
as ``...distribution of mass values, with a spread, $\delta m$,
related to the mean life $\delta \tau (=1/\lambda)$, by uncertainty
relation $\delta m \delta \tau \sim 1$'' (see Introduction in Ref.
\cite{Matth}). In the Refs. \cite{Greenb, Lich}, UP is described by
the so-called asymptotic free field as the state with indefinite
(not fixed) mass. The hypothesis of continuous (smeared, indefinite)
mass of UP was developed in a series of works, where a quantum field
model of UP was presented (see, for instance, review articles
\cite{Kuksa1, Kuksa2} and references therein). In this approach, the
physical values of the mass and width are related to the parameters
of continuous mass distribution. It should be noted that the
definition of the spectral function does not follow from the first
principles. So, it was constructed in phenomenological ways and has
a different form in above mentioned papers. Moreover, the spectral
function is sensible to the "tails" of distribution.

In this work, we consider the structure of the propagators in the
framework of the spectral-representation approach and with account
of the Dyson procedure. As was noted early, Dyson summation is not
well-defined at peak range, while the spectral approach can not be
applied far from the peak. So, we have used the information which
follows from the both approaches in the domains of their
validity. We suppose that the propagator of scalar UP in
spectral representation coincides with the BW one in the
intersection of their domains of definition. Using this assumption we
define the spectral function of boson UP and apply it for the case
of the vector UP's propagator. We show, that this strategy strictly
leads to the propagators which have the structure of the modified
Breit-Wigner ones (\ref{1.2}) under the condition that the spectral
function is defined for both positive and negative values of its parameter (see
the second section).

The paper is organized as follows. In the second section we present
the principal elements of the approach and analyze the general
structure of the scalar propagator. The expressions for vector and
fermionic propagators are derived in the third and fourth sections
respectively. Some aspects of Lagrangian formalism with unstable
field ingredient and respective conclusions concerning the physical
status of the results are made in the fifth and last sections.

\section{Propagator of scalar unstable particle}

The structure of propagator for the case of scalar UP can be
represented in the following convolution form:
\begin{equation}\label{2.1}
D(q)=
i\int_{s_0}^{\infty}\,\frac{\rho(m^2)\,dm^2}{q^2-m^2+i\epsilon}=
i\int_{s_0}^{\infty}D_0(q^2,m^2)\rho(m^2)dm^2,
\end{equation}
where $\rho(m^2)$ is spectral function of the parameter $m^2$,
$D_0(q^2,m^2)$ is ``bare'' scalar propagator and the limit of
integration $s_0$ will be determined further. Note, the symbolic
expression (\ref{2.1}) has different explicit form in various
approaches. One can get a traditional Lehmann-K\"{a}ll\'{e}n
representation for the case
$\rho(m^2)=\delta(M^2-m^2)+\rho_{LK}(m^2)$, where $\rho_{LK}(m^2)=0$
below the threshold $m^2<4M^2$. In the framework of the asymptotic
free field approaches (indefinite mass) \cite{Greenb, Lich} or the
model with continuous mass \cite{Kuksa1, Kuksa2} the expression
(\ref{2.1}) can be derived directly. In these cases, the field
function of scalar UP can be represented in the following
convolution form:
\begin{equation}\label{2.2}
\phi(x)=\frac{1}{(2\pi)^{3/2}}\int\,
\int\phi(\mathbf{p},m^2)e^{ipx}d\mathbf{p}\,\omega(m^2)\,dm^2,
\end{equation}
where $p=(\mathbf{p},p^0)$, $\phi(\mathbf{p},m^2)$ is defined in
standard way at fixed mass $p^2=m^2$ and $\omega(m^2)$ is model
weight function. Note, the value $m$ is not a conventional observed
mass of UP. It is continuous mass parameter which cuts out
three-dimensional surface in the four momentum space according to
equality $p^2=m^2$. The canonical commutation relations contain an
additional delta-function $\delta(m^2-m'{}^{2})$. Starting from the
standard definition of the Green's function $D(q)=i\int dx\,
exp\,(-iqx)\,\langle 0|\hat{T}\phi(x)\phi(0)|0\rangle$, where
$\phi(x)$ is defined by (\ref{2.2}), by straightforward calculations
we get convolution representation of the model propagator
(\ref{2.1}), where $\rho(m^2)=|\omega(m^2)|^2$.

The principal problem of the approach under consideration is to
define the spectral function $\rho(m^2)$. In this connection we
should note the general peculiarity of the spectral approaches. From
the expression (\ref{2.1}) it follows the problem with threshold
value of the spectral parameter $s_0$. Propagator of UP near the
threshold $q^2\approx s_0$ contains divergent at $q^2\to
s_0$ contributions which are compensated far
from the threshold. This threshold effect is explicitly described with
the help of the known integration rule
\begin{equation}\label{2.3}
\int_{a}^{b}\frac{f(x)\,dx}{x\pm i\epsilon}=\mp i\pi f(0)+
\mathcal{P}\int_{a}^{b}\frac{f(x)}{x}dx,
\end{equation}
which follows from the Sokhotski-Plemelj formula when $x=0\in
(a,b)$. In Eq.~(\ref{2.3}) $\mathcal{P}\int$ stands for the Caushy principal
value of the integral.
The threshold effect is caused by the pole at $q^2=s_0$ and will be
described further in more detail.

Here, we consider the special case of the spectral function for
scalar UP in the assumption that the scalar propagator has a
conventional BW form:
\begin{equation}\label{2.4}
D^{BW}(q)=\frac{1}{q^2-M^2+iM\Gamma}.
\end{equation}
In Eq.(\ref{2.4}), we use $q$-independent scheme of the width
insertion and omit general factor $i$ for simplicity. Scalar
propagator in this form can be derived by Dyson summation too, if we
use renormalization conditions
$\mathsf{M}_0^2=\mathsf{M}^2-\Re\Pi(\mathsf{M}^2),\,
Z^{-1}=1+\Re\Pi^{'}(\mathsf{M}^2)$ and unitary condition
$Z\Im\Pi(M^2)=-\sqrt{M^2}\Gamma(M^2)$ \cite{Willenb,LopezC2} (note,
the definitions of the $\Pi(\mathsf{M}^2)$ in these references have
different sign). Taking into account the above mentioned
peculiarities of summation procedure, we use the expression
(\ref{2.4}) as phenomenological postulate which was verified
by the experiments with good accuracy. Note also, that the expression (\ref{2.4})
has the asme status in the framework of CMS which
does not rely on the Dyson procedure. Starting from the BW
expression (\ref{2.4}) for scalar propagator, we shall define the
corresponding spectral function $\rho(m^2)$ and apply it to derive
the expressions for the propagators of vector and spinor UP.

To define $\rho(m^2)$ we rewrite Eq.(\ref{2.1}) with the help of the
integration rule (\ref{2.3}). Then, Eq.(\ref{2.1}) takes the form:
\begin{equation}\label{2.5}
D(q)=-i\pi\rho(q^2)+\mathcal{P}\int\frac{\rho(m^2)}{q^2-m^2}\,dm^2\,.
\end{equation}
The condition $D(q)=D^{BW}(q)$ leads to a following equalities:
\begin{align}\label{2.6}
\Im D(q)={}&-\pi\rho(q^2)=\frac{-M\Gamma}{(q^2-M^2)^2+M^2\Gamma^2},\notag\\
\Re D(q)={}&\mathcal{P}\int\frac{\rho(m^2)\,dm^2}
{q^2-m^2}=\frac{q^2-M^2}{(q^2-M^2)^2+M^2\Gamma^2},
\end{align}
where the first equalities follow from (\ref{2.5}) and the second
ones from (\ref{2.4}). From the upper equality in (\ref{2.6}) it
follows that
\begin{equation}\label{2.7}
\rho(m^2)=\frac{1}{\pi} \frac{M\Gamma}{(m^2-M^2)^2+M^2\Gamma^2}.
\end{equation}
Thus, the condition $\Im{D(q)}=\Im{D^{BW}(q)}$ uniquely defines the
form of the function $\rho(m^2)$ for the case under consideration
($q$-independent $M$ and $\Gamma$). In Ref.\cite{Jacob} the
definition of the function $\rho(m^2)$ was given in close
analogy with above consideration and was finished at this stage.
Here, we take into consideration the lower equality of (\ref{2.6})
which gives an additional information about the limits of
integration. By straightforward calculation we can check that the
lower equality of Eq.(\ref{2.6}) and normalization of the function
(\ref{2.7}) are fulfilled exactly if $(-\infty <m^2 <\infty)$.
Inserting the expression (\ref{2.7}) into the lower equality of
(\ref{2.6}) we get:
\begin{align}\label{2.8}
\Re
D(q)={}&\frac{M\Gamma}{\pi}\mathcal{P}\int_{s_0}^{\infty}\frac{dm^2}{(q^2-m^2)[(m^2-M^2)^2+M^2\Gamma^2]}\notag\\
    ={}&\frac{M\Gamma}{\pi Q}[\frac{\pi p}{2d}+\frac{1}{2}\ln
    \frac{(q^2-s_0)^2}{(q^2-s_0)^2-p(q^2-s_0)+Q} +\frac{p}{d}\arctan
    \frac{M^2-s_0}{M\Gamma}],
\end{align}
where $Q=(q^2-M^2)^2+M^2\Gamma^2$, $p=2(q^2-M^2)$, and $d=2M\Gamma$.
From the expression (\ref{2.8}) it follows that the lower equality
(\ref{2.6}) is exact when $s_0=-\infty$. So, the parameter $m^2$ can
take a negative value and we have to consider an analytic
continuation of the traditional spectral approach. On the other
hand, the expression (\ref{2.8}) explicitly describes above
mentioned threshold effect. The second term gives a logarithmic
singularity at $q^2\to s_0$, where the point $q^2=s_0$ is cut out by
the integration rule (\ref{2.3}). The nature of this singularity
differs from the one of the conventional threshold singularity,
which takes place in OMS renormalization scheme \cite{Kniehl,Lucio}.
In the framework of the traditional Lehmann-like spectral approach,
threshold effect is absent from the very beginning, because of
$\rho(m^2\leq s_0)=0$. At the same time, BW form of scalar
propagator can not be reproduced exactly in this case. In our
approach, $\rho(m^2)>0$ on the whole real axis, the threshold effect
disappears at $s_0\to -\infty$, and BW form is reproduced exactly.
It should be noted that this effect has rather un-physical
artificial nature (see, also, comments in Section 5).

Let us consider the theoretical status of the result and possible
consequences of the presence of negative mass parameter $m^2<0$ in
the integral representations (\ref{2.1}) and (\ref{2.2}). The
condition $D(q)=D^{BW}(q)$ defines the integral equation which
contains the unknown function $\rho(m^2)$ and was solved exactly.
The form of the spectral function is defined strictly by the choice
of the dressed scalar propagator
as input condition. It should be noted that appearance of the
negative component can be caused by the choice of the BW
approximation. However, we do not know correct (exact) expression
for the input propagator and evaluate the error of approximation. In
the framework of the approaches with continuous mass the negative
component $m^2<0$ leads to the states
with imaginary mass parameters which are usually interpreted as
tachyons. The problem of existence of tachyons is under
considerable discussion in the last decades. The main attention is
paid to the principal problems such as violation of causality,
tachyon vacuum, and radiation instability. It should be noted, that
these problems are related to UP as an observable object with fixed
imaginary mass. In the framework of the effective model
\cite{Kuksa2} UP is described by the positive mass square $M^2$ and
we have no tachyons in the set of physical states.

Now we evaluate the contribution of the negative component. The
spectral function $\rho(m^2)$ is normalized and can be interpreted
as the probability density of parameter $m^2$. So, the probability
of the negative component is:
\begin{equation}\label{2.9}
P(m^2<0)=\int_{-\infty}^0 \rho(m^2; M, \Gamma)\,dm^2 \approx
\frac{\Gamma}{\pi M},\,\,\,(\frac{\Gamma}{M}<<1)
\end{equation}
From (\ref{2.9}) it follows that this probability is proportional to
the factor $\Gamma/M$ which defines the finite-width effects in the
processes with UP's participation. This fact can lead to an
interesting possible conclusions: tachyon instability is intrinsic
property of UP; it can be interpreted as the cause of unstable
particle decay. Now, we evaluate the relative contribution of the
negative component to the full propagator which we define as the
relation:
\begin{equation}\label{2.10}
\epsilon(q^2)=\frac{\int_{-\infty}^{0}\,
D_0(q^2,m^2)\,\rho(m^2)\,dm^2}{\int_{-\infty}^{+\infty}\,
D_0(q^2,m^2)\,\rho(m^2)\,dm^2}.
\end{equation}
In the expression (\ref{2.10}) denominator is full BW propagator
(\ref{2.4}) and the integration in numerator can be performed
directly at $q^2>0$. As a result, we get:
\begin{equation}\label{2.11}
\epsilon(q^2;M,\Gamma)=\frac{1}{\pi}\,\frac{\Gamma
M}{q^2-M^2-i\Gamma M}\,[\frac{1}{2}\ln
\frac{q^4}{M^2(M^2+\Gamma^2)}+\pi\frac{q^2-M^2}{\Gamma M}],
\end{equation}
where we used the approximation $\arctan (M/\Gamma)\approx \pi/2$ in
the second term. From (\ref{2.11}) it follows strong $q^2$
-dependence of the relative contribution $\epsilon(q^2;M,\Gamma)$.
In particular, at the peak range $\epsilon(M^2;\Gamma,M)\approx
-i\Gamma^2/2\pi M^2$, at $q^2>>M^2$ it has asymptotic
$\epsilon(q^2)\to 1$ and at $q^2<<M^2$ from (\ref{2.10}) it follows:
\begin{equation}\label{2.12}
\epsilon(q^2;M,\Gamma)=\frac{\Gamma}{\pi M}\,[\frac{1}{2}\ln
\frac{M^2(M^2+\Gamma^2)}{q^4}+\pi\frac{M}{\Gamma}],
\end{equation}
So, at small $q^2$ the value $\epsilon(q^2;M,\Gamma)$ is large and
we can not cut off the negative component. At $q^2<0$, an upper
integral in (\ref{2.10}) can be calculated with the help of the
integration rule (\ref{2.3}) and calculation gives the same effect.
This effect is a direct consequence of the integration rule
(\ref{2.3}) and connected with the above described threshold effect.
Thus, the account of the negative component is essential for the
case of deep virtual states of UP, that is far from the peak range.
This conclusion arises in any quantum field model with spectral
representation of the propagator in the form (\ref{2.1}) and scalar
propagator in BW form (\ref{2.4}). It should be noted, however, that
the status of the above given evaluations and conclusions crucially
depends on the difference between the BW approximation and exact
finite propagator (which, unfortunately, is unknown).

\section{Propagator of vector unstable particles}

In this section, the result (\ref{2.7}) is applied to determine the
structure of vector UP's propagator. Here, we suggest that the
function $\rho(m^2)$ for boson UP (scalar and vector) is universal.
Such a suggestion is in accordance with the mass redefinition scheme
$M^2=M^2_0+\Re{\Pi(M)}$ and relation $M\Gamma=\Im{\Pi(M)}$ for the
case of both scalar and vector UP. First of all, we demonstrate the
consistency of the expression (\ref{2.7}) for $\rho(m^2)$ and input
condition $D(q)=D^{BW}(q)$ with the help of contour integration,
which will be used in further considerations. According to
Eqs.~(\ref{2.1}) and (\ref{2.7}) the propagator of scalar UP can be
written as follows:
\begin{equation}\label{3.1}
D(q)=\frac{1}{\pi}\int_{-\infty}^{+\infty}\frac{M\Gamma\,dm^2}
{(q^2-m^2+i\epsilon)[(m^2-M^2)^2+M^2\Gamma^2]}.
\end{equation}
This expression can be represented in the form
\begin{equation}\label{3.2}
D(q)=-\frac{1}{\pi}
\int_{-\infty}^{+\infty}\frac{M\Gamma\,dm^2}{(m^2-z_0)(m^2-z_+)(m^2-z_-)},
\end{equation}
where $z_0=q^2+i\epsilon$ and $z_{\pm}=M^2\pm iM\Gamma$. Analytic
continuation of the integrand function in (\ref{3.2}), where $m^2\to
z$, has three poles $z_0,z_+,z_-$ in the complex plane. It decreases
as $1/|z|^2$ for $|z|\to\infty$, that is, it satisfies the condition
$|f(z)|<N/|z|^{1+\delta}$ for $|z|>R_0$, where $N$ and $\delta$ are
positive numbers and $R_0\to\infty$. So, we can apply the method of
contour integration and rearrange $D(q)$ as follows:
\begin{align}\label{3.3}
D(q)=&\mp\frac{M\Gamma}{\pi}\oint_{C_{\pm}}\frac{dz}{(z-z_0)(z-z_+)(z-z_-)}\notag\\
=&2\pi i \sum_{k} \operatorname{Res}(f(z),z_k).
\end{align}
In Eqs. (\ref{3.3}) $k$ is number of the poles,
$\operatorname{Res}(f(z),z_k)$ is the residue at the pole $z_k$ and
$C_{\pm}$ is a contour in the upper ($C_+$) or lower ($C_-$) half of
the complex $z$-plane. The simplest way to perform the integration
is to go along the contour $C_-$ which encloses only one pole $z_-$:
\begin{align}\label{3.4}
D(q)={}&\frac{M\Gamma}{\pi}\oint_{C_-}\frac{dz}{(z-z_-)}\frac{1}{(z-z_+)(z-z_0)}\notag\\
={}&\frac{2i
M\Gamma}{(z_--z_+)(z_--z_0)}=\frac{1}{q^2-M^2+iM\Gamma}.
\end{align}
In Eqs.~(\ref{3.4}) we have used the equality $z_--z_+=-2iM\Gamma$.
One can check that the same result follows from the integration
along the contour $C_+$.

Thus, UP can be described in the framework of two different
hierarchical levels---``fundamental'' level, by the integral
representations (\ref{2.1}), (\ref{2.2})  and phenomenological one,
by the effective theory after integrating out unobservable mass
parameter $m^2$ according to (\ref{3.4}). In the framework of the
effective theory, UP is described by the observed physical values
$M$ and $\Gamma$, which can always be defined as a positive
quantity. So, at this phenomenological level UP has no explicit
tachyonic content which could lead to the above mentioned problems.
Instead, we get the term $iM\Gamma$ which describes the instability
in a traditional way.

To define the structure of vector propagator, we assume that the
spectral function $\rho(m^2)$ is the same as for a scalar UP. Using
the standard vector propagator for a free vector particle with a
fixed mass, we get:
\begin{equation}\label{3.6}
D_{\mu\nu}(q)=\frac{1}{\pi}\int_{-\infty}^{+\infty}\frac{-g_{\mu\nu}+q_{\mu}q_{\nu}/(m^2-i\epsilon)}
{q^2-m^2+i\epsilon}\frac{M\Gamma\,dm^2}{[m^2-M^2]^2+M^2\Gamma^2}.
\end{equation}
In the term $q_{\mu}q_{\nu}/(m^2-i\epsilon)$ we use the same rule of
going around pole as in the denominator $q^2-(m^2-i\epsilon)$. The
integral in Eq.~(\ref{3.6}) can be evaluated with the help of the
formula (\ref{2.3}), however, it is easier to do it using the method
of contour integration. The integration along the lower contour
$C_-$ gives:
\begin{align}\label{3.7}
D_{\mu\nu}(q)={}&-\frac{M\Gamma}{\pi}\oint_{C_-}\frac{(g_{\mu\nu}-q_{\mu}q_{\nu}/(z-i\epsilon))\,dz}
{(z-z_-)(z-z_+)(z-z_0)}\notag\\={}&-2iM\Gamma\frac{g_{\mu\nu}-q_{\mu}q_{\nu}/(z_-)}{(z_--z_+)(z_--z_0)}=
\frac{-g_{\mu\nu}+q_{\mu}q_{\nu}/(M^2-iM\Gamma)}{q^2-M^2+iM\Gamma}\,.
\end{align}
One can check that the integration along the upper contour $C_+$ or
with the help of the formula (\ref{2.4}) leads to the same result.
The expression (\ref{3.7}) coincides with the well-known expression
for modified BW propagator (\ref{1.2}) which satisfies to
electromagnetic Ward identity \cite{LopezC}.

We should note, that both the scalar and vector propagators of UP
can be represented in the form with universal complex mass squared:
\begin{equation}\label{3.8}
D(q)=\frac{1}{q^2-M^2_P};\qquad
D_{\mu\nu}(q)=\frac{-g_{\mu\nu}+q_{\mu}q_{\nu}/M^2_P}{q^2-M^2_P}\,,
\end{equation}
where the structure $M^2_P=M^2-iM\Gamma$ usually is called as
complex-mass definition. This definition is the base element of the
so-called complex-mass scheme of calculation \cite{Denner1,Denner2}.
The dressed propagator of a bosonic UP can be formally obtained from
the ``free" propagator by the substitution $M_0^2-i\epsilon
\longrightarrow M^2-iM\Gamma$. So, the infinitesimal value
$\epsilon$, which formally defines the rule of going around pole in
bare propagator, is an analog of the infinitesimal width of the
intermediate state in the framework of the model approach.

\section{Propagator of spinor unstable particles}

The propagator of a free fermion can be represented in two
equivalent forms:
\begin{equation}\label{4.1}
\hat{D}(q)=\frac{1}{\hat{q}-m+i\epsilon}=\frac{\hat{q}+m-i\epsilon}{q^2-(m-i\epsilon)^2}.
\end{equation}
According to the above mentioned formal rule for constructing the
dressed propagator, we have to make the substitution $m-i\epsilon\to
M-i\Gamma/2$. Then, the dressed propagator of the spinor UP takes
the form (\ref{1.2}). Now, we show that the expression (\ref{1.2})
can be derived in a more systematic way with the help of the
integral representation:
\begin{equation}\label{4.2}
\hat{D}(q)=\int\frac{\hat{q}+m-i\epsilon}{q^2-(m-i\epsilon)^2}\,\rho(m)\,dm\,,
\end{equation}
where the integration range is not defined yet. The spectral
function $\rho(m)$ for fermions differs from the bosonic one,
because of another parametrization $M(q)=M_0+\Re\Sigma(q)$ and
$\Gamma(q)=\Im\Sigma(q)$. The spectral function  for the case of the
spinor UP is as follows:
\begin{equation}\label{4.3}
\rho(m)=\frac{1}{\pi}\,\,\frac{\Gamma/2}{[m-M]^2+\Gamma^2/4}
=\frac{1}{\pi}\frac{\Gamma/2}{(m-M_-)(m-M_+)},
\end{equation}
where $M_{\pm}=M\pm i\Gamma/2$. The main difference between boson
and spinor cases is a presence of the linear term $m$ instead the
quadratic one $m^2$, which is defined at the whole real axis $m^2
\in (-\infty,+\infty)$. Here, we consider a straightforward relation
between the bosonic parameter range and spinor one. Thus, we have
two intervals, namely $(+ i\infty , i0;\,\,\, 0 , \infty)$ and $(-
i\infty , i0;\,\,\, 0 , \infty)$ for the value $m$. In the method of
contour integration the signs $\pm$ correspond to integration along
the contours $C_{\pm}$, which enclose the first or fourth quadrants
of the complex plane. Then, from Eqs.~(\ref{4.2}) and (\ref{4.3}) it
follows:
\begin{equation}\label{4.4}
\hat{D}_{\pm}(q)=\pm\frac{\Gamma}{2\pi}\int_{C_{\pm}}\frac{(\hat{q}+z)\,dz}
{(z^2-z^2_0)(z-z_-)(z-z_+)}\,,
\end{equation}
where $z^2_0=q^2+i\epsilon$, $z_{\pm}=M_{\pm}$ and $C_{\pm}$ are the
above described contours. By simple and straightforward calculations
we can see that the correct result follows from the integration
along the contour $C_-$, while the integration along the $C_+$ leads
to non-physical result. This is likely caused by the presence of the
branch point $z^2_0$ in the first quadrant. From Eq.~(\ref{4.4}) it
follows:
\begin{align}\label{4.5}
\hat{D}_{-}(q)=&-\frac{\Gamma}{2\pi}\int_{C_{-}}\frac{dz}{z-z_-}\,\frac{\hat{q}+z}
{(z^2-z^2_0)(z-z_+)}\notag\\=&-i\Gamma(q)\frac{\hat{q}+z_-}{(z^2_--z^2_0)(z_--z_+)}
=\frac{\hat{q}+M-i\Gamma/2}{q^2-(M-i\Gamma/2)^2}.
\end{align}
The last expression in (\ref{4.5}) coincides with the corresponding
expression in (\ref{1.2}). The spinor complex mass definition
differs from the bosonic one, however, it has similar pole-type
complex structure. Then, the pole definition of the mass and width
of the spinor UP is $M_P=M_{\rho}-i\Gamma_{\rho}/2$ in our
consideration.

\section{Complex-mass scheme in effective theory}

The main result of the previous two sections is the
modified BW expressions (\ref{1.2}) derived in analytical way. We have
showed also the connection between scalar, vector, and spinor
propagator with the help of the spectral approach. Note, the
analogous connection takes place for the case of free fields, that
is for the stable particle approximation \cite{Bogolub}:
\begin{align}\label{5.1}
D_{ik}(x)=&(g_{ik}+\frac{1}{M^2}\frac{\partial^2}{\partial^i\partial^k})D(x)=\frac{1}{(2\pi)^4}\int
\frac{(g_{ik}-\frac{k_i k_k}{M^2})e^{-ikx}}{k^2-M^2}dk,\notag\\
 \hat{D}(x)=&(i\hat{\partial}+M)D(x)=\frac{1}{(2\pi)^4}\int\frac{\hat{k}+M}{k^2-M^2}e^{-ikx}\,dk.
\end{align}
In (\ref{5.1}) the value $M^2=M^2_0-i\epsilon$ and $D(x)$ is scalar
casual function, i.e. propagator in coordinate representation for
free scalar field $\Phi(x)$ which satisfies to Klein-Gordon
equation:
\begin{equation}\label{5.2}
(\partial_k\partial^k-M^2)\Phi(x)=0.
\end{equation}

We have derived just the same expressions for the case of dressed
propagators in the momentum representation, that is for complex mass
$M= M_P$, which should be inserted into Lagrangian (vertexes,
$\sin^2\theta_W$, etc.) and into motion equation (\ref{5.2}). In
such a way, we have come to the so-called complex-mass scheme of
calculation \cite{Denner1,Denner2,Denner3}, which is realized at
Lagrangian level of effective theory. It should be noted, that free
casual functions $D_{ik}(x)$ and $\hat{D}(x)$, which are defined by
Eqs.(\ref{5.1}), in analogy with scalar one can be represented as
vacuum expectation value of chronological field operator product
\cite{Bogolub}:
\begin{equation}\label{5.3}
D_{ik}(x-y)=i<0|
T(\phi_i(x)\phi_k(y))|0>\,;\,\,\,\hat{D}(x-y)=i<0|
T(\psi(x)\bar{\psi}(y))|0>.
\end{equation}
As it was shown in Section 2, the scalar field function can be
redefined according to (\ref{2.2}) with weight function
$\omega(m^2)=\pm\sqrt{\rho(m^2)}$ which leads to BW propagator. The
same redefinition can be done for the case of vector and spinor UP,
which provides the validity of Eqs.(\ref{5.3}). So, we have
effective theory which has some principal properties of fundamental
quantum field theory. As was noted in the third section, UP can be
described at two hierarchical levels, fundamental and
phenomenological one (after integrating out unobservable mass
parameter $m^2$). In the case under consideration, the
phenomenological approach corresponds to the effective theory, which
can be represented at lowest order by the Lagrangian with complex
masses. The effective field function, which describes UP, is defined
in a correspondence with the conventional definition of Green's
function.

The problems of renormalization procedure in the effective theory
arise at next-to-leading order. To date there is no fully
established treatment of UP within perturbation theory, although
many solutions have been proposed \cite{Denner3}. For instance, the
unitarity in scalar field theories was studied within the framework
of the CMS \cite{Denner3} at one-loop approximation. Evidently, the
effective field function of UP is formed by self-energy contribution
at fundamental level and contains corresponding information about
mass and width of UP. In the case under consideration, this
information is included into spectral function $\rho(m^2;M,\Gamma)$.
So, we have to avoid double counting of self-energy contribution in
the calculations at loop level of the effective theory. Just this
contribution stipulates the divergence of renormalization constant
at threshold $M\to 2M_V$, where the vectors are in the self-energy
loop \cite{Kniehl2}. Thus, we do not deal with the conventional TS
in the framework of the effective theory. As was shown in the second
section, the problem of TS appears at the leading order, and its
nature differs from the conventional one. It should be noted,
however, that mathematically strict status of our consideration is
due to appearing of the negative spectral component which has not
clear physical meaning.

\section{Conclusions}

 The definitions of the mass and width of UP, as a rule, are closely
connected with the construction of the dressed propagators. It was
underlined in this work, that traditional approaches, which are
based on Dyson procedure and spectral representation, have formally
crucial peculiarities. We have considered the structure of the
propagators of UP in the phenomenological approach which is based on
the spectral representation. The spectral function describes the
distribution of continuous (indefinite, smeared) mass parameter,
contains a principal information concerning UP, and defines a
spectral structure of the propagators.

In this work, we have analyzed a special case of the spectral
function which follows from the matching the model and standard
scalar BW propagator. This function contains the parameters $M,\,
\Gamma$ and mass variable $m^2$ prove to be in the interval
$(-\infty,+\infty)$. So, the variable $m$ can be imaginary, however,
such states have no explicit physical content. It was shown that
contribution of the negative component to the full propagator is
significant for the deep virtual states. In the framework of this
approach we get vector and spinor propagators with the well-known
modified BW structure. This structure provides the gauge invariant
description and explicitly leads to the complex-mass definition.
The $q$-dependence of the UP mass and width can be introduced into
the function $\rho(m^2;\Gamma(q),M(q))$ without the loss of the
generality. We formulated some problematical aspects of the
propagators construction which require an additional analysis.

\textbf{Conflict of Interests}\\The author declare that there is no
conflict of interests regarding the publication of this paper.

\textbf{Acknowledgment}\\ The work has been supported by Southern
Federal University grant No. 213.01-2014/013-VG.

\end{document}